\documentclass[UTF8]{article}
\usepackage{amsmath}
\usepackage{amssymb}
\usepackage{graphicx}
\usepackage{geometry}
\usepackage{multirow}
\usepackage{makecell}
\usepackage{epstopdf}
\usepackage{verbatim}
\usepackage{ulem}
\usepackage{color}
\usepackage{float}
\usepackage{CJK}
\newcommand{\spur}[1]{\not\! #1 \,}

\begin{document}
\title{\bfseries Constraints on $Wtb$ anomalous coupling with  $ B\to X_{s}\gamma $ and $B_{s}\to\mu^{+}\mu^{-}$\footnote{Supported by National
Natural Science Foundation of China (NFSC No. 11775012)}}\date{}\maketitle
\begin{center}
\author{Zhao-Hua Xiong$^{1}$, Lian Zhou$^{1}$}\\
$^{1}$ Institute of Theoretical Physics, Beijing University of Technology, Beijing 100124, China
\end{center}
\begin{abstract}
We calculate the amplitudes of $b\to s$ transition in extension of
the Standard Model with $Wtb$ anomalous couplings. We found that i)
there exist the Ward identity violating terms in effective vertix of
$b\to s\gamma$. The terms, which come from the tensor parts of $Wtb$
anomalies, and can be canceled exactly by introducing corresponding
$Wtb\gamma$ interactions; ii) $Br(B_{s} \to \mu^{+}\mu^{-})$
provides unique information on $\delta v_L$ which is set to zero in
top decay experiments, and stringent bounds on $v_R,\ g_L$ by
$Br(B\to X_s\gamma)$  are obtained.

\noindent\textbf{Keywords:}  $Wtb$ Anomalous Coupling;  Wilson
Coefficients; Rare B Precesses
\index{Keywords}

{\bf PACS}: 12.60.Fr, 14.40.Nd, 14.65.Ha

\end{abstract}

\section{Introduction}
The standard model (SM) is a very successful model in describing
physics below the Fermi scale and is in good agreement with the most
experiment data, especially the last particle predicted in the SM
has been discovered at LHC\cite{Aad:2012tfa,PW}. And yet, the
completion of the SM particle spectrum does not mean the completion
of the SM itself until all the  SM interaction forces could be
firmly measured. It is universally accepted that there must be
something more than what has been so brilliantly conceived and
verified.

One important issue is searching the new physics (NP) signatures
directly, for example, probing $Wtb$ couplings in
top quark decay. With its mass around the electroweak symmetry
breaking (EWSB) scale, the top quark is believed to play an
important role to connect the SM and new physics. A first glance at the top
decay has been given through the study of $W$ helicity fractions and
related observables\cite{Wh-ATLAS}. Production cross sections have
been measured both for $t\bar t$
pairs\cite{Khachatryan:2010ez,Aad:2010ey} and for single top
quarks\cite{st-CMS}. The lastest analysis  for $Wtb$ anomalous
couplings by top decay at LHC has presented in \cite{2016twb}.
Another important issue is searching the new physics signatures
indirectly, i.e., probing $Wtb$ anomalous couplings at some of which derive
from flavour  changing neutral-current processes (FCNC), such as
rare B processes.  The new B factories, BELLE and LHCb, are
providing us now with more and more high statistics data, which will
carry FCNC tests to the next precision level. Measurements of rare B
meson decays such as $B \to X_s \gamma$, $B_s\to \mu^+\mu^-$
\cite{LHCb12}, are likely to provide sensitive tests of the SM
\cite{Bernardo14,Grzadkowski08}. These rare B decays have been
studied extremely at the leading logarithm order\cite{BLOSM},
 high order in the SM\cite{BHOSM} and various new physics models \cite{NewphysB1,
NewphysB2,NewphysB3}.

Currently, most works on anomalous $Wtb$ couplings are probing NP
directly in  processes of top quark production
/decay\cite{Bernardo14,Gao15}, only a few authors have studied
virtual effect of the anomalous $Wtb$ couplings on some rare decays
through loop-level\cite{Grzadkowski08,Jong08}. In
Ref.\cite{Grzadkowski08}, the authors calculated contribution of the
anomalous couplings to $b\to s\gamma$ and presented some constraints
on four parameters, but they did not presented a proof that the
effective vertex of $b\to s\gamma$ satisfies the Ward identity.
Ref.\cite{Jong08} only considered the $V+A$ contributions of $Wts$
to $B_s-\bar{B_s}$ mixing.  The contribution is just  proportional
to the anomalous coupling squared, which is hard to present a strong
constraints on the parameters.

In this paper, we will evolute the effect of the $Wtb$ anomalous
couplings on the rare B decays. We will present  a detailed one-loop
correction for the $b\to s\gamma$, including contributions from all
anomalous couplings related to the third generation quarks such as
$\bar{t}t\gamma$, $\bar{b}b\gamma$, as well as $Wtb\gamma$. We will
prove that  with effect of new $Wtb\gamma$ anomalous couplings taken
into account,  the   Ward identity violating terms in the effective
vertex with $Wtb$ anomalies cancel exactly, and this is independent
of the unitarity of the quark mixing matrix.  Then we obtain the
Wilson coefficients $C_i$ ($i=7$ to $10$)  for branching ratios of
$B\to X_s\gamma$ and $B_s\to \mu^+\mu^-$. We also present some
stringent constraints on the four anomalous couplings.

The work  is organized as follows. In Section 2, we  will render a
detailed one loop calculation for $b\to s \gamma^{(*)}$, proving the
effective vertex satisfies the Ward identity.  And then we obtain
the Wilson coefficients from the effective vertices of $b\to
s\gamma$ and $b\to s \ell^+\ell^-$. In Section 3, we express the
branching ratios of $B\to X_s\gamma$ and $B_s\to \mu^+\mu^-$ in
terms of the $Wtb$ anomalous couplings,  and present some numerical
 bounds for the anomalous couplings. Some inputs
are presented in the Appendix.

\section{Amplitudes of $b\to s $ transition}
\subsection{Effective vertices of $b\to s\gamma$}
 Before going to detailed calculation, let us start with the most
 general $Wtb$ vertex:
 \begin{equation}\label{wtbl}
 \mathcal{L}_{Wtb}=-\frac{gV_{tb}^{*}}{\sqrt{2}}\bar{b}[\gamma_{\mu}(v_{L}L+v_{R}R)+\frac{i\sigma_{\mu\nu}k^{\nu}}
 {m_{W}^{2}}(g_{L}L+g_{R}R)]tW^{-\mu}
 \end{equation}
 where $v_{L}=1+\delta v_{L}$. In the SM, $\delta v_{L}$, $v_{R}$,
 $g_{L}$, $g_{R}$ are zero.  $L,\ R =(1\mp\gamma_{5})/2$, and
 $k$ is the W boson momentum. In the phenomenology, new physics
 effects can be parameterized in terms of an effective field theory
 approach.  Then the  effective vertex $\Gamma_{\mu}$ in amplitude
 $A(b\to s\gamma)$ has form as:
 \begin{equation}
 \Gamma_{\mu}=e\dfrac{g^2}{32\pi^{2}}V_{tb}V_{ts}^{*}
[\gamma_{\mu}LF_{0}+\dfrac{m_{b}^{2}}{m_{W}^{2}}\gamma_{\mu}LF_{1}
 +\frac{q^{2}\gamma_{\mu}-\spur{q}q_{\mu}}{m_{W}^{2}}LF_{2}+\dfrac{im_{b}\sigma_{\mu\nu}q^{\nu}R}{m_{W}^{2}}LF_{3}] \label{ffactor}
 \end{equation}
 where $q$ is the photon momentum. It is clear that the first two terms
 donate the Ward identity violating terms. Note differing from the top decay which occurs
 at tree level, $b\to s$ processes occur at loop level. In model we research
 the couplings of three up-type quark with $W$ boson are not universal, only top quark in
 loop  contribution to factors $F_j$. Thus, the constants and
 divergent terms in $F_{j}$, which will disappear  due to CKM
 unitarity in the SM, may remain. The divergence will be assumed canceled by
 heavy particles in the model. Furthermore,  whether the Ward identity for  $b\to sV$ ($V=g,\ \gamma)$ vertex are guaranteed
 needs to be checked.
\begin{table}[!h]
    \caption{Factors $F_0^i$ for $b\to s\gamma$ with $B=B(m_{t},\
        m_{W})$, $\tilde{B}=\tilde{B}(m_W, m_W)$,
        $\hat{B}=\hat{B}(m_{t},m_{W})$, $\bar{B}=\bar{B}(m_t, m_t)$ and
        $\bar{C}=\bar{C}(m_W,m_t,m_t)$,\ $\tilde{C}=\tilde{C}(m_t,m_W,m_W)$}
\begin{center}
\begin{tabular}{l|r|l}
\hline\hline
$v_{L}$ &$i=1$&
$-\frac{1-\delta_{t}}{2}+2B_0^0-3B_1^0-\frac{1}{m_W^2}[\frac{B_{22}^0}{2}
+A\left(m_{t}\right)]$\\
&2&$\frac{3}{2}-\frac{\delta_{t}}{2}\left(1+\bar{C}_{24}^0\right)
+2m_{t}^2\bar{C}_0^{0}+\frac{\bar{B}^{0}_{22}}{m_W^2}$\\
&3& $\delta_{t}-\frac{3\tilde{B}_{22}^{0}}{2m_{W}^{2}}+\frac{3}{2}(2-\delta_{t})
\tilde{C}_{24}^{0}$\\
\hline
$g_{R}$& 1 &$-1+3B_1^0$\\
& 2 &$1-\frac{3}{2}\bar{C}_{24}^0$\\
&3 &$\frac{1-\delta_{t}}{2}-\frac{3}{2}\tilde{C}_{24}^{0}
+\frac{3\tilde{B}_{22}^{0}}{4m_{W}^{2}}$\\
&4 &$-\dfrac{3-\delta_{t}}{2}+3\bar{B}_{0}-\frac{3}{4}\frac{\bar{B}_{22}^{0}}{m_{W}^{2}}
$\\
\hline\hline
\end{tabular}
\label{tab1}
\end{center}
\end{table}

 It is well known, at matching scale $\mu_W$ the factors $F_j$ are functions of
 $\delta_t=m_t^2/m_W^2$ and can be written as follows:
 \begin{eqnarray}
 F_j=v_L F_{j}^{v_L}
 +v_R\frac{m_t}{m_b}F_{j}^{v_R}+g_L\frac{m_W}{m_b} F_{j}^{g_L}
 +g_R\frac{m_t}{m_W}F_{j}^{g_R}.
 \label{ffactor1}
 \end{eqnarray}
The task left to us is deriving these factors, then obtain the
Wilson coefficients. We use unitary gauge and  take naive dimension
renormalization (NDR)
 scheme in calculation.  The factors $F_j$ receive contributions from
Feynman diagrams of 1) self-energy; 2) vertex with one W in loop; 3)
vertex with two $W$
 bosons; 4) vertex with  $ Wtb\gamma$  couplings and 5) vertex with
 $tt\gamma$ anomalies, as displayed in Fig. 1.
 We express each
 type Feynman diagram  contributions to $F_j$, denoted by $F_j^i$ in
 terms of loop functions $A,\ B,\ C$ and list them in Table
 \ref{tab1} to Table \ref{tab4}.  Then $F_j$ are obtained as
 \begin{equation}
 F_j =e_dF_j^1+e_uF_j^2+F_j^3+F_j^4+F_j^5,
 \end{equation}
 where $e_u=2/3, \ e_d=-1/3$ stand for  charge numbers  of quarks.
To trace our calculation easily, we present  defination for the
loop factions as follows:
\begin{align}
&B_\mu(p;m_1,m_2)=p_\mu B_1, \notag\\
&C_{\mu}(p,p^\prime;m_1,m_2,m_3)=C_{11}p_{\mu}+C_{12}p_{\mu}^{\prime},\notag\\
&C_{\mu\nu}(p,p^{\prime};m_1,m_2,m_3)=C_{21}p_{\mu}p_{\nu}+C_{22}p_{\mu}^{\prime}p_{\nu}^{\prime}+C_{23}(p_{\mu}^{\prime}p_{\nu}+p_{\mu}p_{\nu}^{\prime})+\frac{g^{\mu\nu}}{4}C_{24}
\end{align}
In B physics the loop functions can be expanded order by order as
\begin{align}
&B(p;m_1,m_2)=B^0+\frac{p^2}{m_W^2}B^1+\cdots, \notag\\
&C(p_1,p_2;m_1,m_2,m_3)=C^0+\frac{p_1^2}{m_W^2}C^{1,1}
+\frac{2p_1p_2}{m_W^2}C^{1,2}+\frac{p_2^2}{m_W^2}C^{1,3}+\cdots,
\end{align}
with functions $B^n$, $C^{n,i}$  being independent of momenta. The
definations and corresponding expansions can be found in Ref.
\cite{Gunion86}.
\begin{table}[!h]
    \caption{Factors $F_1^i$ for $b\to s\gamma$}
    \begin{center}
        \begin{tabular}{l|r|l}
            \hline\hline
            $v_{L}$ &$ i= 1 $&$-\frac{1}{6}+2B_0^1-3B_1^1+B_{21}^0-\frac{1}{2}\frac{B_{22}^1}{m_W^2}$\\
            &2&$\bar{B}^{0}_1-\bar{B}^{0}_0-\frac{\delta_{t}}{2}(\bar{C}_{24}^{1,1}
            +\bar{C}_{24}^{1,2})+2m_{t}^2\left(\bar{C}_{0}^{1,1}
            +\bar{C}_{0}^{1,2}\right)+
            +m_W^2(3\bar{C}^{0}_{23}-2\bar{C}^0_{11})$\\
            &3&$\tilde{B}_{0}^{0}-m_W^2[(4-3\delta_{t})\tilde{C}_{11}^{0}
            -(12-3\delta_{t})\tilde{C}_{23}^{0}]-\frac{1}{2}
            \tilde{C}_{24}^{0}+\frac{3(2-\delta_{t})}{2}(\tilde{C}_{24}^{1,1}
            +\tilde{C}_{24}^{1,2})$\\
            \hline
            $v_{R}$
            &1&$\frac{1}{6}-4B_0^1+\frac{B_{22}^1+B_{21}^0}{m_W^2}$ \\
            &2&$-\frac{1}{2}-\bar{B}^{0}_0+\bar{C}_{24}^0+(6\bar{C}_{11}^{0}
            -4\bar{C}_{0}^{0})m_W^2$\\
            &3&$-4\tilde{C}_{24}^{0}+4\tilde{C}_0^{0}-(8-2\delta_{t})
            \tilde{C}_{11}^{0}$\\
            \hline
            $g_{L}$& 1 &$-3\left(B_1^{0}-B_{21}^{0}-\frac{1}{2}\right)
            +\frac{3B_{22}^1}{m_W^2}$\\
            &2&$-2+m_W^2[\bar{C}_0^{0}-2(1+2\delta_{t})\bar{C}_{11}^{0}]
            +\frac{3}{2}\bar{C}_{24}^0$\\
            &3&$1+\frac{\delta_{t}(\hat{B}_{0}^{0}- \hat{B}_{1}^{0})}{2}-\tilde{C}_{24}^{0}
            +m_{t}^{2}(6\tilde{C}_{11}^{0}-3\tilde{C}_{0}^{0})$\\
            &4&$-\bar{B}_{0}+\frac{\bar{B}_{1}}{2}-\frac{A(m_{t})}{2m_{W}^{2}}+\frac{\bar{B}_{22}^{0}}{4m_{W}^{2}}$\\
            \hline
            $g_{R}$& 1 &$3B_1^{1}$\\
            & 2 &$3m_W^2(\bar{C}_{11}^{0}-3\bar{C}_{23}^{0})-\frac{3}{2}(\bar{C}_{24}^{1,1}+\bar{C}_{24}^{1,2})$ \\
            &3&$-3m_W^2(\tilde{C}_{0}^{0}-3\tilde{C}_{11}^{0}
            +3\tilde{C}_{23}^{0})-\frac{3}{2}(\tilde{C}_{24}^{1,1}+\tilde{C}_{24}^{1,2})$ \\
            \hline\hline
        \end{tabular}
        \label{tab2}
    \end{center}
\end{table}

It is time to check our results by Ward identity of
$b\to sg$ and $b\to s \gamma$, which implies $F_0$ and $F_1$ in
(\ref{ffactor}) should be zero. The decay $b\to sg$ is governed by
operator $O_{8}$ in next Subsection and also contribution to $b\to
s\gamma$ as QCD correction included. Considering the strong
interactions with quarks are independent of quark charges, using the
analytic formulae presented in Table \ref{tab1} to  Table
\ref{tab4}, we find $F_0^1+F_0^2= F_1^1+F_1^2=0$ as expected.
However, this is not the case in $b\to s\gamma$.  The effective
vertex of $b\to s\gamma$ receives contributions not only from
diagrams to $b\to s g$ with $g$ replaced by $\gamma$, also diagram
with two W bosons in loop. We find that
\begin{align}
e_dF_0^1+e_uF_0^2+F_0^3&=0\label{Lorentzinv2}
\end{align}
is not satisfied for $g_R$ term, and
\begin{align}
e_dF_1^1+e_uF_1^2+F_1^3&=0\label{Lorentzinv3}
\end{align}
is not satisfied for the $g_L$ term. This implies that without the
new anomalous interactions introduced as extracting $Wtb$ anomalous
couplings in top decay,  the effective vertex of $b\to s\gamma$ does
not satisfy the Ward identity. It is obvious, the source of the new
anomalous couplings should the same as $g_L$ and $g_R$. We expect
that new introduced contributions can cancel part contributions to
$g_L,\ g_R$ terms.

In this work, we follow Ref.\cite{Grzadkowski08} where $Wtb\gamma$
anomalous couplings are always along with the tensor parts of $Wtb$
interactions. They have the origin from dimension-six operators:
\begin{eqnarray}
&&Q_{LRt}=\bar{q}_{L}\sigma^{\mu\nu}\tau^{a}t_{R}\tilde{\phi}
W_{\mu\nu}^{a}+h.c.,\nonumber\\
&& Q_{LRb}=\bar{q\prime}_{L}\sigma^{\mu\nu}\tau^{a}b_{R}\phi
W_{\mu\nu}^{a}+h.c.
\end{eqnarray}
Here $\phi$ denotes the Higgs doublet,
$\tilde{\phi}=i\tau^{2}\phi^{*} $, and $W_{\mu\nu}^{a}$ is the
field strength and ${q}_{L}=(t_{L}, V_{tb}^{*}b_{L})^{T}$,
${q}_{L}^{\prime}=(V_{tb}t_{L}, b_{L})^{T}$. Note that for
consistency, in this work we do not consider $Wts$, $Wcb,\ Wub$
anomalies, but take $tt\gamma$ and $bb\gamma$ anomalies into
account. The additional anomalous interactions are given by
\begin{equation}
\delta\mathcal{L}=-\dfrac{ge}{\sqrt{2}m_{W}}iV_{tb}^{*}\bar{t}\sigma^{\mu\nu}
(g_{R}L+g_{L}R)bW_{\nu}^{+}\epsilon_{\mu}+\dfrac{1}{2}e\bar{t}
\dfrac{i\sigma^{\mu\nu}q_{\nu}}{m_{W}}g_{R}t\epsilon_{\mu}-
\dfrac{1}{2}e\bar{b}\dfrac{i\sigma^{\mu\nu}q_{\nu}}{m_{W}}g_{L}b\epsilon_{\mu},
\end{equation}
where $\epsilon$ is the photon polarization vector. Our treatment
is some different from Ref.\cite{Grzadkowski08}. We will discuss
more in next subsection.

\begin{figure}[H]\label{fig1}
	\begin{minipage}[t]{0.5\linewidth}
		\centering
		\includegraphics[width=2.2in]{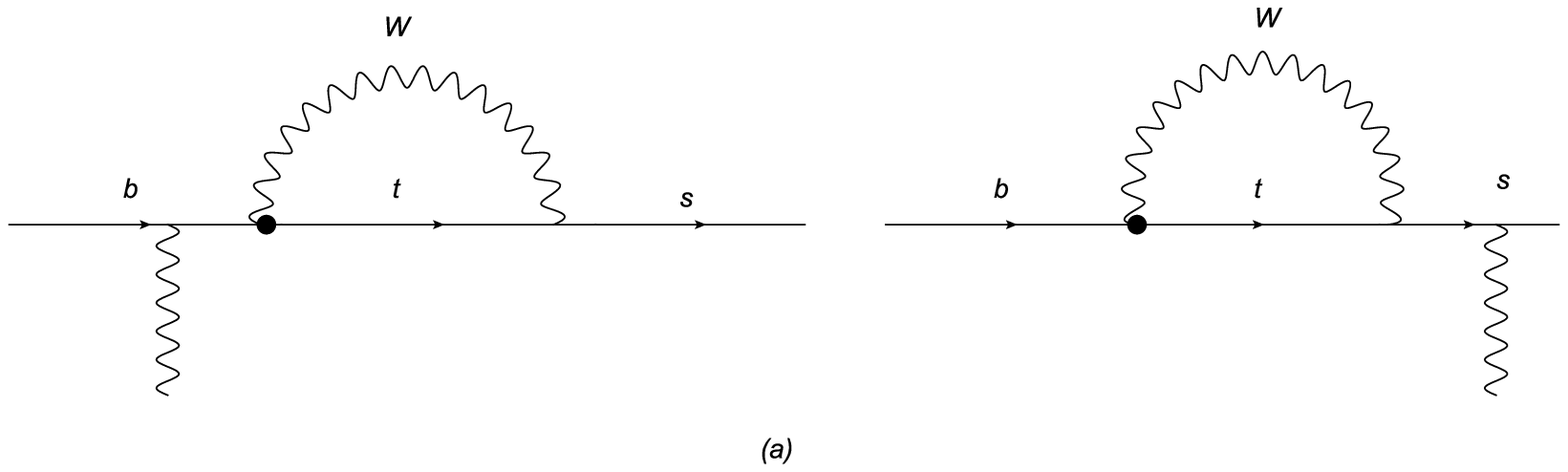}
		\label{fig1a}
	\end{minipage}
	\begin{minipage}[t]{0.5\linewidth}
		\centering
		\includegraphics[width=2.2in]{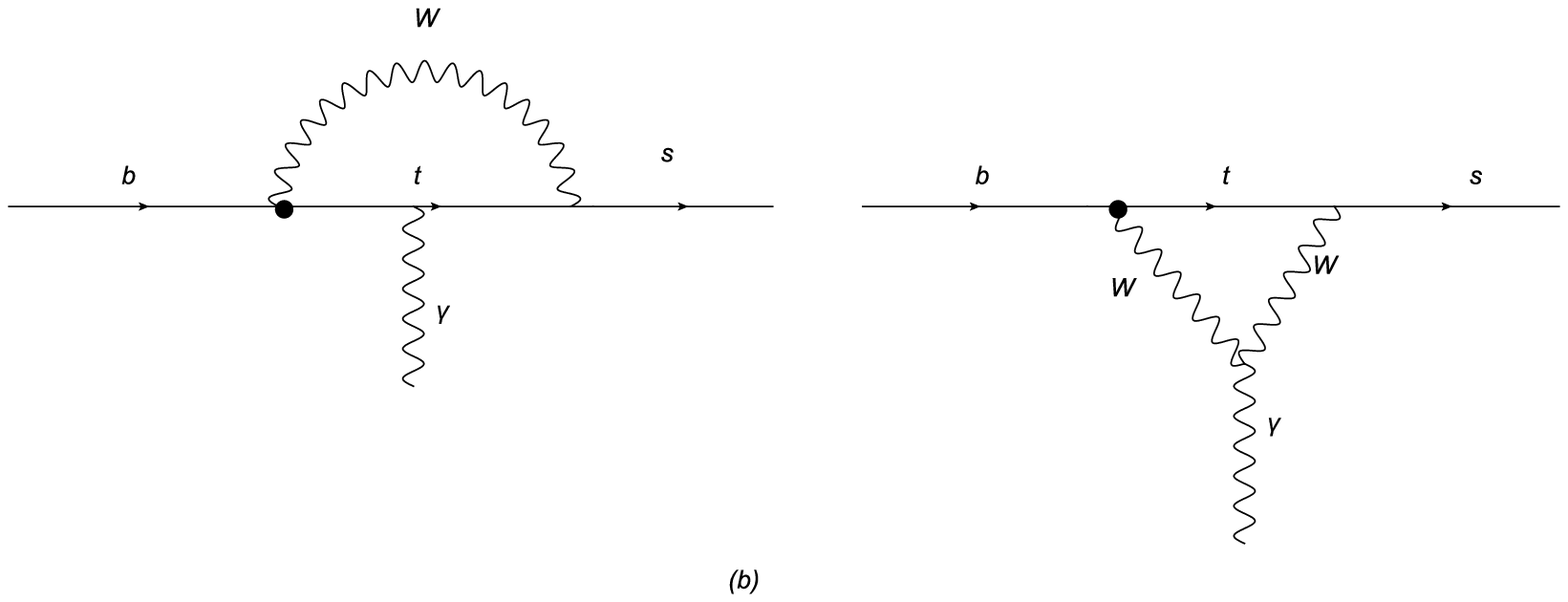}
		\label{fig1b}
	\end{minipage}
\begin{minipage}[t]{0.5\linewidth}
	\centering
	\includegraphics[width=2.2in]{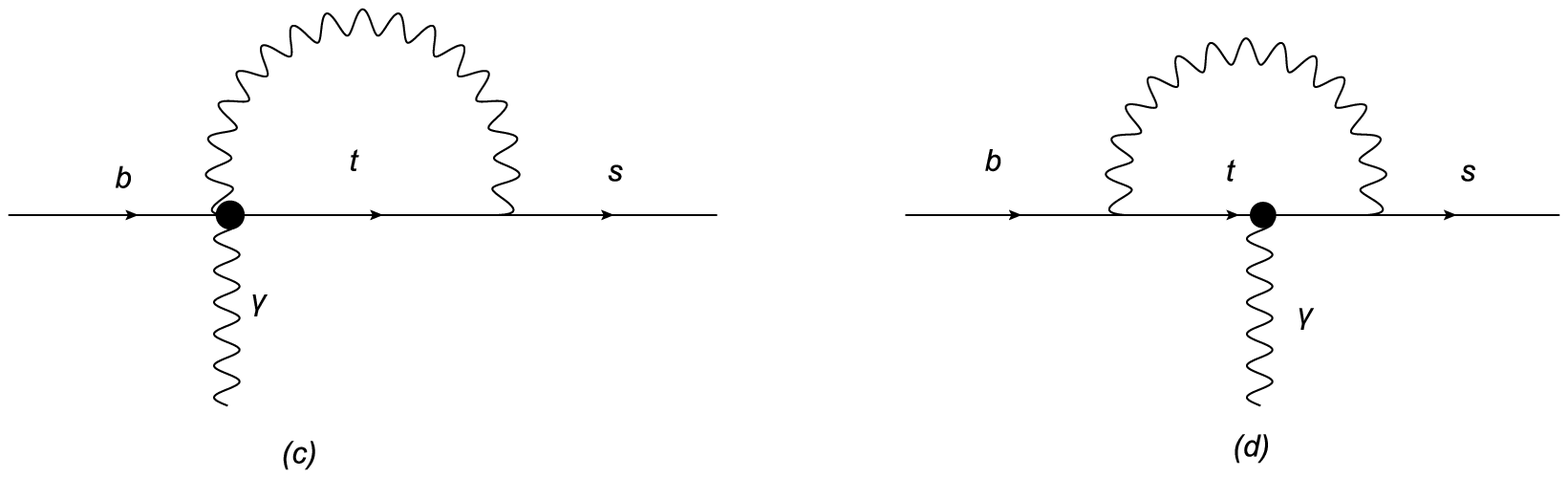}
	\label{fig1c}
\end{minipage}
\begin{minipage}[t]{0.5\linewidth}
	\centering
	\includegraphics[width=2.2in]{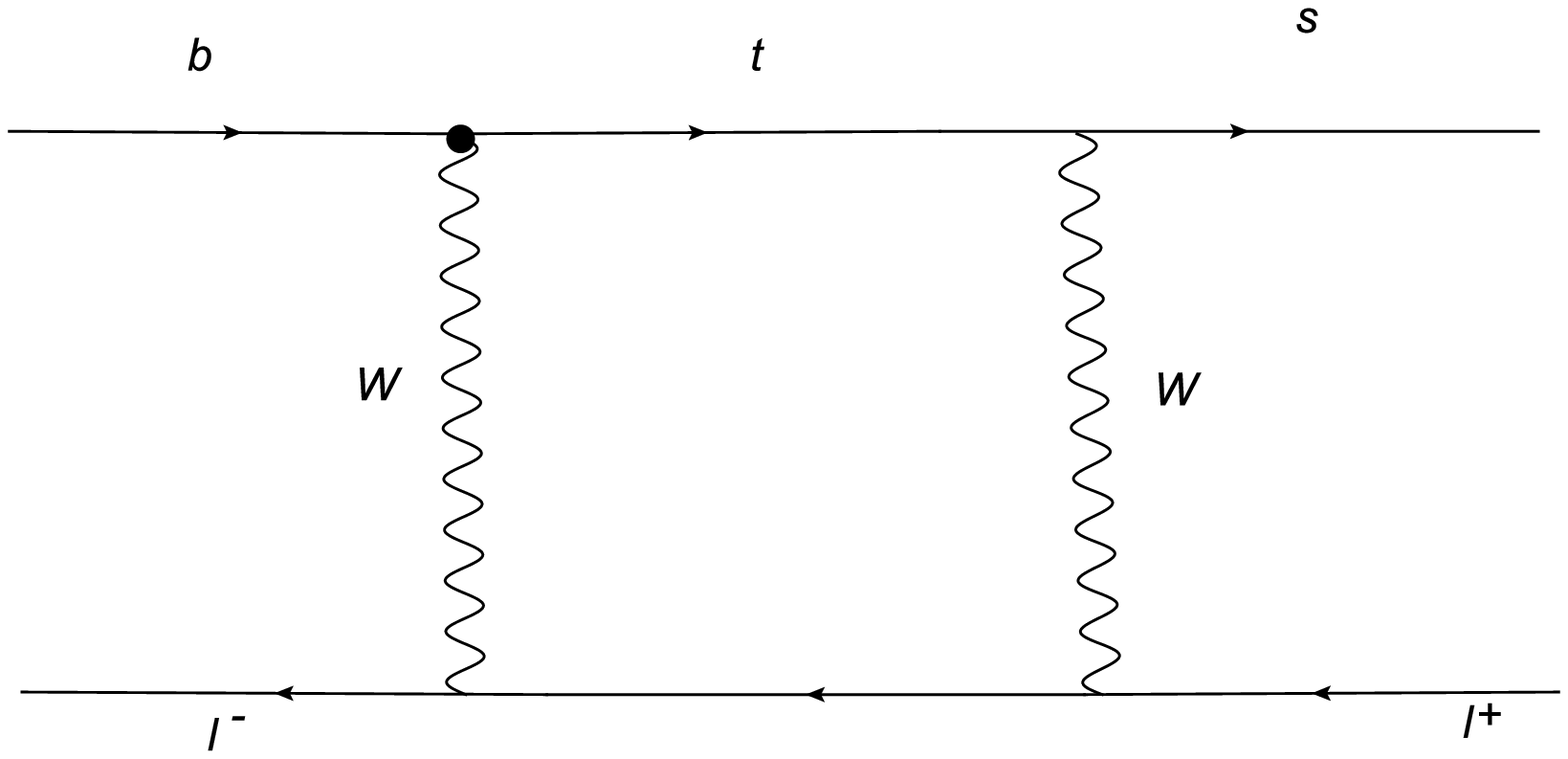}
\end{minipage}
        \caption{The Feynman diagrams of $b\to s$ transition. (a) Self-energy ;
            (b) Vertices with one loop with one and two $W$ bosons;
            (c) Vertex with $Wtb\gamma$ anomalous couplings; 
            (d) Vertex with $tt\gamma$ anomalous couplings and Box diagram}
\end{figure}
With the Feynman rules given in above, we find  the $bb\gamma$
anomalous has no contribution to $b\to s\gamma$. The Feynman
diagrams of $Wtb\gamma$ and $tt\gamma$ contribution to $b\to
s\gamma$ are plotted in Fig. 1. The corresponding factors $F_j$ are
also listed in Table \ref{tab1} to Table \ref{tab4}. We find
$F_{0}^{g_R}=F_{1}^{g_L}=0$. Thus, the Ward identity in $b\to
s\gamma$ requires that the tensor parts of $Wtb$ anomalies should be
accompanied  with the $Wtb\gamma$ anomalies. Although the $tt\gamma$
anomalous couplings are not related to the Ward identity, for
consistency its contribution to $g_L$ part of $b\to s\gamma$ also
should been included.
\begin{table}[!h]
    \caption{Factors $F_2^i$ for $b\to s\gamma$}
    \begin{center}
        \begin{tabular}{l|r|l}
            \hline\hline
            $v_{L}$ &$i=2 $&$\frac{2}{3}+\bar{B}^{0}_{21}+\bar{B}^{0}_0-\bar{B}^{0}_1
            +\frac{\bar{B}_{22}^{1}}{m_{W}^{2}}+m_W^2[2\bar{C}_0^{0}-2\bar{C}_{11}^{0}+(1+\delta_{t})\bar{C}_{23}^{0}
            -2\delta_{t} \bar{C}_0^{1,2}]-\frac{1}{2}\bar{C}_{24}^0
            +\frac{\delta_{t}}{2}\bar{C}_{24}^{1,2}$\\
            &3&$-\frac{1}{3}+\tilde{B}^{0}_0+2\tilde{B}^{0}_1-2\tilde{B}^{0}_{21}
            -\frac{1}{m_{W}^{2}}(\frac{\tilde{B}_{22}^{0}}{4}
            +\frac{3}{2}\tilde{B}_{22}^{1})+m_W^2[2\delta_{t}\tilde{C}_{0}^{0}
            +2(2-\delta_{t})\tilde{C}_{11}^{0}-2(1-\delta_{t})\tilde{C}_{23}^{0}]$\\
            & &$-\frac{\delta}{4}\tilde{C}_{24}^0-\frac{3(2-\delta_{t})}{2}
            \tilde{C}_{24}^{1,2}$\\
            \hline
            $g_{R}$& 2 &$m_W^2\bar{C}_{11}^{0}+\frac{3}{2}\bar{C}_{24}^{1,2}$\\
            & 3 &$\frac{1}{2}+m_{W}^{2}(\tilde{C}_{0}-4\tilde{C}_{11})+\frac{3}{2}\tilde{C}_{24}^{1,2}-\frac{3}{4}\tilde{C}_{24}^{0}$\\
            &5&$-\frac{1}{2}\bar{B}_{0}^{0}-m_{W}^{2}(2\bar{C}_{0}-\bar{C}_{11}^{0})+\frac{\bar{C}_{24}^{0}}{2}$\\
            \hline\hline
        \end{tabular}
        \label{tab3}
    \end{center}
\end{table}

\subsection{Effective vertex of $b\to s\ell^+\ell^-$}
In this subsection, we calculate the effective vertex of $b\to
s\ell^+\ell^-$. The vertex receives contribute for \ $F_{3}$ which
is related to the operator $O_7$, $F_2$ which is related to the
operator $O_9$ in vertex of $b\to s\gamma$, i.e., off-shell
$\gamma$- penguin, and the Z-penguin, $F^Z$, which can be obtained
by replacement  $\gamma$ with Z boson in $b\to s\gamma$ diagrams.
Using the relation $F_0^1=-F_0^2=F_0^3+F_0^4$ we obtain
\begin{align}
F^{Z,v_L}&=\bar{C}_{0}^{0}m_{t}^{2}-\frac{\delta_{t}}{4}(\bar{C}_{24}^{0}-1),
\nonumber\\
F^{Z,g_R}&=-\frac{3}{4}\bar{C}_{24}^{0}+\frac{5}{4}.
\end{align}
The effective vertex of $b\to s\ell^+\ell^-$ also receives
contribution from box diagrams, as displayed in Fig. 1. The
contributions denoted by $F^{B}$  read:
\begin{align}
F^{B,v_L}&=m_{W}^{2}\tilde{C}_{0}^{0}
-\frac{7}{4}\tilde{C}_{24}^{0}+\dfrac{\tilde{B}_{22}^{0}}{4m_{W}^{2}}+1,\nonumber\\
F^{B,g_R}&=\frac{3}{2}\tilde{C}_{24}^{0}-1.
\end{align}

At the end of this Section,  we obtain the Wilson coefficients of
operators at matching scale by comparing the amplitudes of $b\to
s\gamma $ and $b\to s \ell^+\ell^-$ with  the effective Hamiltonian:
\begin{equation}
{\cal  H}^{eff}(b\to
s)=-\dfrac{4G_{F}}{\sqrt{2}}V_{tb}V_{ts}^{*}\sum_{i=1}^{10}C_{i}(\mu)O_{i}.
\label{hamiltion}
\end{equation}
The operators in (\ref{hamiltion}) related to  the factors $F_j$
directly are defined as follows:
\begin{align}
O_{7}=&\dfrac{e}{16\pi^{2}}m_{b}\bar{s}_{i}\sigma^{\mu\nu}Rb_{i}F_{\mu\nu}\notag\\
O_{8}=&\frac{g_{s}}{16\pi^{2}}\bar{s}_{i}\sigma^{\mu\nu}RT_{ij}^{a}b_{j}G_{\mu\nu}^{a},\notag\\
O_{9}=&\frac{e^{2}}{16\pi^{2}}(\bar{s}\gamma^{\mu}Lb)(\bar{\ell}\gamma_{\mu}\ell)\notag
\\
O_{10}=&\frac{e^{2}}{16\pi^{2}}(\bar{s}\gamma^{\mu}Lb)
(\bar{\ell}\gamma_{\mu}\gamma_{5}\ell),
\label{o910}
\end{align}
where $F_{\mu\nu}$ and $G_{\mu\nu}$ are the electromagnetic tensor
and strong tensor, respectively, and $T^a$ stand for the  $SU(3)$
generators. Note  coefficients of the four quark operators $O_i$
(i=1 to 6) are the the same as those in the SM, other Wilson
coefficients at matching scale are written in terms of factor $F_j$
as:
\begin{align}
C_{7}(\mu_W)&=-\frac{1}{2}F_3, \notag\\
C_8(\mu_W)&=-\frac{1}{2}F_3^2,\notag\\
C_9(\mu_W)&=\frac{F^Z+F^B}{4\sin^2\theta_W}-F^Z+F_2,\notag\\
C_{10}(\mu_W)&=-\frac{F^B+F^Z}{4\sin^2\theta_W}.
\label{Wisloncoeff}
\end{align}
Note in the SM, $C_9({\mu_W})$ has a constant 4/9 which is come from
the correction of the four-quark operators \cite{BLOSM}.

\begin{table}[H]
    \caption{Factors $F_3^i$ for $b\to s\gamma$}
    \begin{center}
        \begin{tabular}{l|r|l}
            \hline\hline
            $v_{L}$ &$i=2$&$\frac{1}{2}+\bar{B}_{1}^{0}+m_W^2[3(2+\delta_{t})C23+2\bar{C}_0^{0}
            -5\bar{C}_{11}^{0}]-\frac{1}{2}\bar{C}_{24}^0$\\
            &3&$\tilde{B}_{1}^{0}+m_{W}^{2}[\delta_{t} \tilde{C}_{0}^{0}
            +(1-2\delta_{t})\tilde{C}_{11}^{0}+3(\delta_{t}+2)\tilde{C}_{23}^{0}]-\frac{\tilde{C}_{24}^{0}}{2}$\\
            \hline $v_{R}$ &2&$-\bar{B}^{0}_0+\bar{C}_{24}^0+(6\bar{C}_{11}^{0}-4\bar{C}_{0}^{0})m_W^2$\\
            &3&$\frac{1}{2}-6m_W^2\tilde{C}_{11}^{0}$\\
            \hline
            $g_{L}$ &2&$-1+\bar{B}_{1}^{0}+m_W^2[2\bar{C}_0^{0}-3(1+\delta_{t})\bar{C}_{11}^{0}]
            +\frac{3}{2}\bar{C}_{24}^0$\\
            &3&$
            -1-\frac{A(m_{W})}{m_{W}^{2}}+2\tilde{B}_{0}^{0}-\frac{\delta_{t}(\hat{B}_{0}^{0}       +\hat{B}_{1}^{0})}{2}+\frac{\tilde{C}_{24}^{0}}{2}+m_{t}^{2}(6\tilde{C}_{11}-\tilde{C}_{0}^{0})$\\
            &4&$-\bar{B}_{0}+\frac{\bar{B}_{1}}{2}-\frac{A_{0}(m_{t})}{2m_{W}^{2}}+\frac{\bar{B}_{22}^{0}}{4m_{W}^{2}}$\\
            \hline
            $g_{R}$&2&$3m_W^2(\bar{C}_{11}^{0}-3\bar{C}_{23}^{0})$\\
            &3&$3m_{W}^{2}(\tilde{C}_{11}^{0}-3\tilde{C}_{23}^{0})$\\
            &5&$\frac{1}{4}-\frac{1}{2}\bar{B}_{0}^{0}-m_{W}^{2}(\bar{C}_{0}-\bar{C}_{11}^{0})+\frac{\bar{C}_{24}^{0}}{4}$\\
            \hline\hline
        \end{tabular}
        \label{tab4}
    \end{center}
\end{table}

For comparison with the SM results, we first present the Wilson
coefficients related to $v_L$ at matching scale for $b\to s\gamma$:
\begin{align}
C_7^{v_L}(\mu_W)&=\dfrac{3\delta_{t}^{3}-2\delta_{t}^{2}}{4(\delta_{t}-1)^{4}}\ln \delta_{t}+\frac{22\delta_{t}^{3}-153\delta_{t}^{2}+159\delta_{t}-46}{72(\delta_{t}-1)^{3}}\notag\\
C_8^{v_L}(\mu_W)&=-\frac{3\delta_{t}^{2}}{4(\delta_{t}-1)^{4}}\ln \delta_{t}+\frac{5\delta_{t}^{3}-9\delta_{t}^{2}+30\delta_{t}-8}{24(\delta_{t}-1)^{3}}
\label{coeffw1}
\end{align}
It is clear that
\begin{align}
C_{7}^{SM}=C_{7}^{v_{L}}-\frac{23}{36},\nonumber\\
C_{8}^{SM}=C_{8}^{v_{L}}-\frac{1}{3},\nonumber
\end{align}
where the constants in $C_{7,8}^{v_L}$  in the SM disappear due to
the CKM unitarity \cite{BLOSM}. Consider the
$C_{7}^{SM}(m_{W})=-0.19$, one can see the constant is very
important in phenomenological analysis.

When $C_{9}$ and $C_{10}$ related to $v_L$ are compared with the SM
values for $b\to s\ell^+\ell^-$, we stress that the factors $F_j$ we
obtained by using unitary gauge, the formulae  for $B=-F^B/4$,
$C=F^Z/4$, $D=-F_2$ used in expressing $C^{SM}_{9,10}$ are not the
same as those calculated in Feynman gauge even the constants deleted
by the CKM unitarity. In this sense, comparing  our calculation with
$B^{SM}$, $C^{SM}$, $D^{SM}$ separately has no meaning since they
are gauge dependent. However, from Eq.(\ref{Wisloncoeff}) one can
see that $C-B=(F^B+F^Z)/4$ and $4C+D=F^Z-F_2$ should be independent
of gauge expect for constants and divergent terms. They are
presented as follows:
\begin{align}
\frac{1}{4}(F^{B,v_L}+F^{Z,v_L})
&=-\frac{3}{4}\log(\frac{\mu_{W}}{m_{W}})-\frac{1-\delta_{t}+5\delta_{t}^{2}}{8(1-\delta_{t})}
+\frac{3\delta_{t}^{2}\ln\delta_{t}}{8(1-\delta_{t})^{2}}\notag\\
&=(C-B)^{SM}-\frac{3}{4}\log(\frac{\mu_{W}}{m_{W}})-\frac{1}{8},\notag \\
F^{Z,v_L}-F_2^{v_L}&=\frac{1}{3}\log(\frac{\mu_{W}}{m_{W}})+\frac{26-534\delta_{t}-363\delta_{t}^{2}+379\delta_{t}^{3}
-54\delta_{t}^{4}}{108(1-\delta_{t})^{3}}
+\frac{8-74\delta_{t}+159\delta_{t}^{2}-90\delta_{t}^{3}}{18(1-\delta_{t})^{4}}\log\delta_{t}\notag\\
&=(4C+D)^{SM}-\frac{8}{3}\log(\frac{\mu_{W}}{m_{W}})+\frac{61}{54}.
\label{coeffw2}
\end{align}

Other Wilson coefficients read:
\begin{align}
C_{7}^{v_{R}}&=-\dfrac{3\delta_{t}^{2}-2\delta_{t}}{2(\delta_{t}-1)^{3}}\ln \delta_{t}+\frac{-5\delta_{t}^{2}+31\delta_{t}-20}{12(\delta_{t}-1)^{2}}\\
C_{8}^{v_{R}}&=\frac{3\delta_{t}}{2(\delta_{t}-1)^{3}}\ln \delta_{t}-\frac{\delta_{t}^{2}+\delta_{t}+4}{4(\delta_{t}-1)^{2}}\\
C_{7}^{g_{L}}&=(\delta_{t}-\frac{7}{3})\log(\frac{\mu_{W}}{m_{W}})+\frac{8-27\delta_{t}+7\delta_{t}^{2}+6\delta_{t}^{3}}{12(1-\delta_{t})^{2}}-\frac{\delta_{t}(2+3\delta_{t}+\delta_{t}^{2}-3\delta_{t}^{3})}{6(1-\delta_{t})^{3}}\log\delta_{t}\\
C_{8}^{g_{L}}&=-2\log(\frac{\mu_{W}}{m_{W}})+\frac{2-9\delta_{t}+\delta_{t}^{2}}{4(1-\delta_{t})^{2}}-\frac{\delta_{t}(1-6\delta_{t}+2\delta_{t}^{2})}{2(1-\delta_{t})^{3}}\log\delta_{t}\\
C_{7}^{g_{R}}&=\frac{1}{4}\log(\frac{\mu_{W}}{m_{W}})+\frac{17-31\delta_{t}+5\delta_{t}^{2}-3\delta_{t}^{3}}{48(1-\delta_{t})^{3}}+\frac{2-2\delta_{t}-3\delta_{t}^{2}+2\delta_{t}^{3}-\delta_{t}^{4}}{8(1-\delta_{t})^{4}}\\
C_{8}^{g_{R}}&=\frac{2+5\delta_{t}-\delta_{t}^{2}}{8(1-\delta_{t})^{3}}+\frac{3\delta_{t}}{4(1-\delta_{t})^{4}}\log\delta_{t}\\
\frac{1}{4}(F^{B,g_R}&+F^{Z,g_R})=\frac{3}{8}\log(\frac{\mu_{W}}{m_{W}})-
\frac{17\delta_{t}+1}{32(1-\delta_{t})}-
\frac{3\delta_{t}(\delta_{t}+2)\log \delta_{t}}{16(1-\delta_{t})^2}\notag\\
F^{Z,g_R}&-F_2^{g_R}=
\frac{34-35\delta_{t}-59\delta_{t}^{2}+54\delta_{t}^{3}}{36(1-\delta_{t})^3}
-\frac{(12\delta_{t}^3-18 \delta_{t}^2+\delta_{t}+4)\log
\delta_{t}}{6(1-\delta_{t})^4}.
\label{coeffw3}
\end{align}

From Eqs.(\ref{coeffw1}) and (\ref{coeffw2}), it is clear that
$C_{7,8}^{v_L,v_R}$ and $C_{8}^{g_R}$ originate from the
ultraviolet-finite diagrams and depend on $\delta_{t}$ only.
However, divergent terms appear  in other Wilson coefficients.
Consequently, $ \ln\frac{\mu_{W}}{m_{W}}$ are left in
these functions. We also compare our formulae of $C_{7,8}$ with Ref.
\cite{Grzadkowski08}, find that there is a 1/2 factor in our
calculation of $C_7^{v_L}$, and $C_8^{v_L}$ has different
expression. And there are no  constants and divergences in
$C_{7,8}^{g_L}$ in Ref. \cite{Grzadkowski08} since the authors
included $Wcb$, $Wub$ anomalous couplings contributions, as
mentioned in last subsection. Further, the non-log term in
$C_8^{g_L}$ has opposite sign with that in  Ref.
\cite{Grzadkowski08}. We confirm our calculation.

\section{Constraints on $Wtb$ anomalous couplings with rare decays}

   With all Wilson  coefficients at $\mu_W$ scale ready, we can express
the branching ratios in terms of the Wilson coefficient, further, of
the anomalous couplings.   Because the SM contributions to $
C_{i}(\mu_{b}) $ and the corresponding operator matrix elements are
mostly real, the linear terms in $\delta C_{i}$, which stem from
SM-NP interference contributions contribute mostly as $Re[\delta
C_{i}] $, we neglect the small contributions of
$Im[C_{i}]$\cite{exploring} and set $\mu_W=m_W$. The $\bar{B} \to
X_{s}\gamma$ branching ratio can be expressed as follows \cite{pa}:
\begin{align}
Br(B\to X_{s}\gamma)&=[3.15\pm 0.23-14.81\delta \tilde{C}_{7}(\mu_{b})+16.68(\delta \tilde{C}_{7}(\mu_{b}))^{2}]
\times 10^{-4}
\end{align}
Using $\delta C_{7}(\mu_{b})=0.627\delta C_{7}(m_{W})$, $\delta
C_{8}(\mu_{b})=0.747\delta C_{8}(m_{W})$ and $ \delta
\tilde{C}_{7}^{eff}=\delta C_{7}+0.24\delta C_{8}$, as well as  the
Wilson coefficients in  Eqs (\ref{coeffw1}) and  (\ref{coeffw2}) we
rewritten the branching as
\begin{align}
Br(B\to X_{s}\gamma)&=[3.15\pm 0.23-3.40\delta v_{L}+339.09v_{R}-78.49g_{L}-1.12g_{R}\notag\\
&+(0.94\delta v_{L}-93.51v_{R}+21.65g_{L}+0.31g_{R})^{2}]\times 10^{-4}.
\label{brbsr}
\end{align}
The branching ratio for $B_s\to \mu^{+}\mu^{-}$ is given by
\cite{exploring}
\begin{align}
Br(B_s\to \mu^{+}\mu^{-})&=1.8525\times 10^{-10}[|-4.3085+\delta C_{10}|^{2}\pm1.7274]\notag\\
&=1.8525\times 10^{-10}[|-4.3085-29.17\delta v_{L}-7.11g_{R}|^{2}\pm 1.7274]
\label{brbmm}
\end{align}

From Eqs (\ref{brbsr}) and (\ref{brbmm}), we can see that
$Br(B_s\to \mu^{+}\mu^{-})$ depends only parameters $\delta v_L$ and
$g_R$  and is  more sensitive to $\delta v_{L}$ which is set to zero
in top decay research \cite{2016twb}. And $Br(B\to X_{s}\gamma)$
depend all four parameters and  will exert more stronger limits on
$v_R$ and $g_L$ than in top decay since enhanced factors $m_t/m_b$,
$m_W/m_b$, as pointed out in Ref. \cite{Grzadkowski08}.

\begin{figure}[H]
\centering
\includegraphics{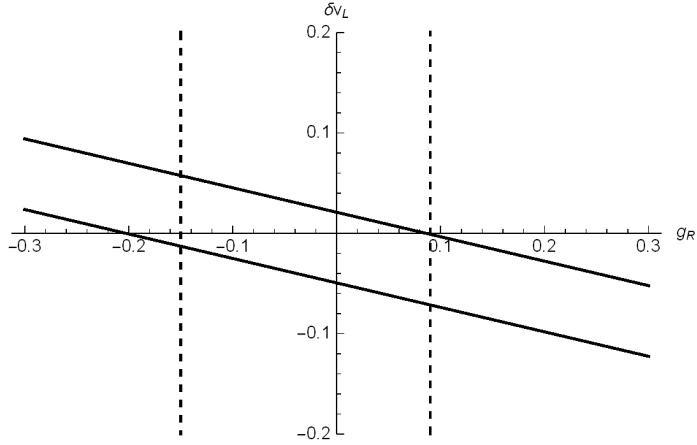}
\label{vlgr}
\caption{$\delta v_L$ as a function of $g_R$ constrained by  $Br(B_s \to \mu^{+}\mu^{-})$
and top decay where $\delta v_L$ is set to zero. The solid lines stand for $2\sigma$ bounds
 for $Br(B_s\to\mu^{+}\mu^{-})$, while the dashed lines from top decay.}
\end{figure}
We first constrain parameters $\delta v_L$ and $g_R$ by $Br(B_s\to
\mu^+\mu^-)$. Using the branching ratio by  experimental
measurements and the SM  prediction, as well as constraints  from
lastest top decay experiment in Appendix, we plot $\delta v_L$ as a
function of $g_R$ in Fig. \ref{vlgr}. We find that $B_{s}\to
\mu^{+}\mu^{-}$ exerts bounds  more tighter restrictions on $\delta
v_{L}$ but slight weaker  on $g_R$ than those at LHC via top decays
with $\delta v_{L}=0$.

\begin{figure}[H]
\centering
\includegraphics[scale=0.8]{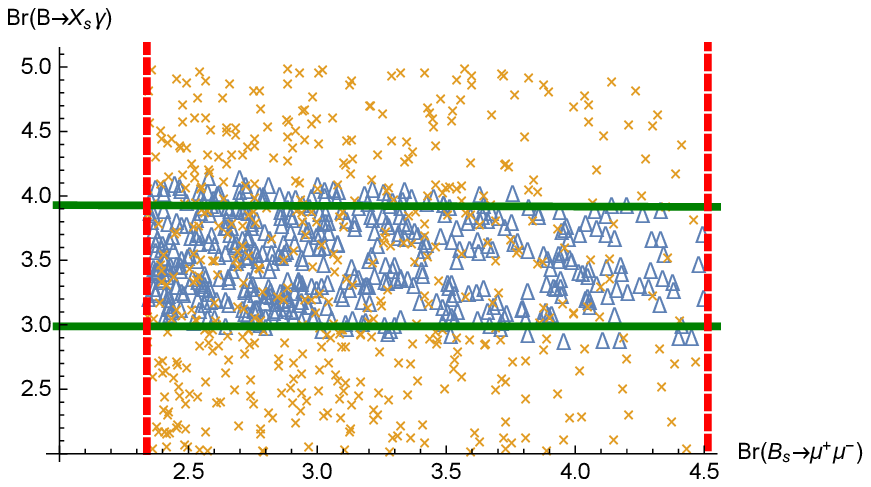}
\caption{The scatter plot of $Br(B\to X_{s}\gamma)$ (in unit $10^{-4}$) as function of
$Br(B_s \to \mu^{+}\mu^{-})$\ (in unit  $10^{-9}$). The parameters  $\delta v_{L}$, $g_{R}$ are restrictioned by
$B_s \mu^{+}\mu^{-}$, and allow region of $g_L,\ v_R$ are obtained  by top decay ( denoted by triangles )
and $Br(B\to X_{s}\gamma)$ in Ref. \cite{Grzadkowski08} (denoted by  crosses). The $2\sigma$ bounds for
$Br(B_s\to  \mu^{+}\mu^{-})$ and $Br(B\to X_{s}\gamma)$ are aslo shown as solid and dashed lines, respectively.}
\label{bsrbsmm}
\end{figure}

Under the limits by $Br(B_{s}\to \mu^{+}\mu^{-})$ and lastest top
decay experiment at LHC \cite{2016twb}, we plot $Br(B\to
X_{s}\gamma)$ as a function of $Br(B_{s}\to \mu^{+}\mu^{-})$ in
Fig.\ \ref{bsrbsmm}.  Since $Br(B_{s}\to X_s\gamma)$ depends on all
four parameters and are more sensitive to $g_L,\ v_R$ which are less
constrained by top decay, leading to very large branching ratio, we
also use the bounds from $Br(B\to X_{s}\gamma)$ presented in Ref.
\cite{Grzadkowski08} for comparison. Note at one loop level,
 $Br(B\to X_{s}\gamma)$ is proportional to $|C_7^{eff}|^2$, the NP
contributions  to the branching ratio may as large as the SM  value.
Although  the anomalous couplings are small, the coefficients of
$v_R,\ g_L$ have large enhancements, as mentioned in Subsection 1.
Thus, we keep up to the anomalous couplings squared in expression of
$Br(B\to X_{s}\gamma)$. Besides, some Wilson coefficients are also
different from those in Ref. \cite{Grzadkowski08}.  From Fig.\
\ref{bsrbsmm} we obtain more stringent constraints on $g_L,\ v_R$
than in top decay and slightly tighter constraints than previous
work \cite{Grzadkowski08}.

Now we would like to summarize our work. We calculated the
amplitudes of $b\to s$ transition in extension of the Standard Model
with $Wtb$ anomalous couplings in unitary gauge. We presented all
anomalous couplings related to the third generation quarks including
anomalous $Wtb $, $ Wtb\gamma$ and $ tt\gamma$ couplings needed in
calculation of rare B decays, and  a detailed one-loop correction
calculation for the $b\to s\gamma^{(*)}$. We found that effective
vertex of $b\to s\gamma$ does not satisfy the Ward identity  without
$Wtb\gamma$ vertex contribution taken into account. This is
independent of the CKM matrix unitarity even the photon is
off-shell. We express branching ratios $B\to X_{s}\gamma$ and
$B_s\to \mu^{+}\mu^{-}$ in terms of the Wilson coefficients and the
anomalous couplings.  The numerical results show that unlike to the
inclusive decays, $Br(B_s\to \mu^{+}\mu^{-})$ depends only on two
anomalous parameters,  it exerts unique  constraints on $\delta v_L$
since $\delta v_L$ is set to zero in top decay at LHC. Under the
constraints, we obtain stringent limits on other parameters by $B\to
X_{s}\gamma$ than obtained at LHC via $t\to W^{+}b$ decay and
previous $B\to X_s\gamma$ analysis. Our work will be useful in
experimental measurements for $Wtb$ anomalous couplings with more
data at high energy colliders and B factories.

\section*{Appendix}
In this Appendix, we present some inputs for numerical analysis.
\section*{1.\  Experimental measurements and SM predictions of the
Branching ratios}
\begin{table}[H]
\begin{center}
\caption{The branching ratios of $B$ decays\cite{pdg2017}}
\begin{tabular}{|c|c|c|}
\hline
Process&$ Br^{ex} $&$ Br^{SM} $\\
\hline
$B\to X_{s}\gamma$&$ (3.43\pm 0.21\pm 0.07)\times 10^{-4}|_{E\gamma>1.6GeV} $&$ (3.15\pm 0.23)\times 10^{-4}|_{E\gamma>1.6GeV} $\\
\hline
$ B_s\to \mu^{+}\mu^{-}$&$ (3.0\pm0.6^{+0.3}_{-0.2})\times 10^{-9} $&$ (3.42\pm 0.54)\times 10^{-9}$\\
\hline
\end{tabular}
\label{expsm}
\end{center}
\end{table}

\subsection*{2.\ Bounds on $Wtb$ anomalous couplings by top decays and $Br(B\to X_s\gamma)$}
\begin{table}[H]
    \begin{center}
        \caption{$95\%$ C.L. limits on the real components
            of the anomalous couplings. These limits were extracted
            from the combination of $W$-boson helicities and single
            top quark production cross section measurements at LHC at
            14 TeV with high luminosity\cite{2016twb} and $Br(B\to X_s\gamma)$
            \cite{Grzadkowski08}}
        \begin{tabular}{c|c|c|c|c}
             \hline
             &$\delta v_L$&$g_{R}$&$g_{L}$&$v_{R}$ \\
    \hline
    Allowed regions & 0&$[-0.07, 0.07]$&$[-0.16, 0.17]$&$[-0.25,0.34]$ \cite{2016twb}\\
    Allowed regions &$ [-0.13,0.03]$&$[-0.0007,0.0025]$&$[-0.0013,0.0004]$&$[-0.15,0.57]$ \cite{Grzadkowski08}\\
    \hline
        \end{tabular}
    \end{center}
    \label{lhc}
\end{table}

\end{document}